\documentstyle[12pt,a4]{article}

\newcommand{\reduce}{{\sc reduce}}
\newcommand{\D}[2]{{\frac{\partial #1}{\partial #2}}}
\newcommand{\Dt}[1]{{\frac{\partial #1}{\partial t}}}
\newcommand{\Dx}[1]{{\frac{\partial #1}{\partial x}}}
\newcommand{\Dy}[1]{{\frac{\partial #1}{\partial y}}}
\newcommand{\Dyy}[1]{{\frac{\partial^2 #1}{\partial y^2}}}
\newcommand{\DD}[2]{{\frac{\partial^2 #1}{\partial #2^2}}}
\newcommand{\Dn}[3]{{\frac{\partial^#3 #1}{\partial #2^#3}}}
\newcommand{\Ord}[1]{{\cal O}\left(#1\right)}

\newcommand{\cG}{{\cal G}}
\newcommand{\cL}{{\cal L}}
\newcommand{\cM}{{\cal M}}
\newcommand{\cR}{{\cal R}}
\newcommand{\cV}{{\cal V}}
\newcommand{\cZ}{{\cal Z}}
\newcommand{\th}{\tilde{\vec h}}
\newcommand{\tv}{\tilde{\vec v}}
\newcommand{\tg}{\tilde{\vec g}}
\newcommand{\pv}{{\vec v}'}
\newcommand{\pg}{{\vec g}'}
\newcommand{\va}{{\vec a}}
\renewcommand{\vec}[1]{{\bf #1}}
\newcommand{\vel}{{\vec q}}

\newcommand{\grad}{\nabla}
\newcommand{\divv}{\nabla\cdot}
\newcommand{\cm}{{\rm cm}}
\newcommand{\half}{\frac{1}{2}}

\begin{document} 
\title{\bf Low-dimensional modelling of dynamics via computer algebra} 
\author{A.J.~Roberts\thanks{Dept.\ Mathematics \& Computing, 
University of Southern Queensland, Toowoomba 4350, AUSTRALIA. 
E-mail: \tt aroberts@usq.edu.au}} 
\maketitle 


\begin{abstract}
I describe a method, particularly suitable to implementation by computer 
algebra, for the derivation of low-dimensional models of dynamical systems.  
The method is systematic and is based upon centre manifold theory.  
Computer code for the algorithm is relatively simple, robust and flexible.  
The method is applied to two examples: one a straightforward pitchfork 
bifurcation, and one being the dynamics of thin fluid films.
\end{abstract}

\section{Introduction}

Deterministic evolution equations, either ordinary differential equations 
or partial differential equations, are used to describe dynamics in the 
physical world.  Often, these equations allow many modes of behaviour that 
are of little physical interest in particular applications, for example 
sound waves are neglected in many applications of fluid mechanics.  The 
essential dynamical behaviour of the system is then determined by the 
evolution of a subset of the possible modes, the ``critical'' modes.  Rapid 
oscillations or heavy damping characterise the modes that need to be 
eliminated from consideration.  If the amplitudes of the modes are viewed 
as co-ordinates in a state space then the dynamical behaviour of the system 
corresponds to motion along some trajectory.  When many of the modes are 
heavily damped, trajectories are rapidly attracted to some low-dimensional 
invariant manifold, which may be parameterised by the amplitudes of the 
critical modes.  This geometric picture is the heart of the application of 
centre manifolds\cite{Carr81,Coullet83} to the rational construction of 
low-dimensional model systems by the elimination of physically 
uninteresting fast modes of behaviour.  Such low-dimensional dynamical models are significantly easier 
to analyse, simulate and understand.

Applications of the techniques have 
ranged over, for example, triple convection \cite{Arneodo85c}, feedback 
control \cite{Boe89}, economic theory \cite{Chiarella90}, shear dispersion
\cite{Mercer90,Mercer94a}, nonlinear oscillations \cite{Shaw93}, beam theory 
\cite{Roberts93}, flow reactors \cite{Balakotaiah92}, and the dynamics of 
thin fluid films \cite{Roberts94c}.  New insights given by the centre 
manifold picture enable one to not only derive the dynamical models, but 
also to provide accurate initial conditions \cite{Roberts89b,Cox93b}, 
boundary conditions \cite{Roberts92c}, and the treatment of forcing 
\cite{Cox91}.

In the application of centre manifold theory, the typical approach 
is to express the model explicitly in terms of asymptotic sums 
\cite[e.g.]{Coullet83,Arneodo85c,Shaw93,Roberts93,Balakotaiah92,Roberts94c,Mei95}.  
To reduce the dynamics onto the centre manifold, one then has to substitute 
the asymptotic sums into the governing equations, reorder the summations, 
rearrange to extract dominant terms, and evaluate the expressions.  While 
perfectly acceptable when done correctly, it leads to formidable working 
which obscures the construction of a model.  Further, such asymptotic 
expansions, in common with the method of multiple scales, reinforce the 
notion that careful balancing of the ``order'' of small effects are 
necessary in the {\em construction} of a model rather than in its {\em 
use\/} in some situation.  Instead, I propose in \S2 an iterative method, 
based upon the residuals of the governing differential equations, for the 
construction of such low-dimensional, dynamical models.  The evaluation of 
the residuals is a routine algebraic task which may be easily done using 
computer algebra, as seen in the examples of \S\S2 and~3, by simply coding 
the governing differential equations; it replaces the whole messy detail of 
the manipulation of asymptotic expansions ( e.g.~\cite[\S5.4]{Coullet83}).  
The aim of this proposed approach is to minimise human time by using a 
novel algorithm which can be simply and reliably implemented in computer 
algebra with relatively small inefficiencies in the use of computer 
resources.  After all: ``It is unworthy of excellent persons to lose hours 
like slaves in the labour of calculation''\ldots Gottfried Wilhelm von 
Leibniz.

The use of centre manifold theory in the construction of low-dimensional 
modes relies on three theorems.  These specifically address dynamical 
systems written in the form:
\begin{equation}
	\begin{array}{rcl}
		\dot{\vec x} & = & A\vec x+\vec f(\vec x,\vec y)\,,  \\
		\dot{\vec y} & = & B\vec y+\vec g(\vec x,\vec y)\,,
	\end{array}
	\label{stddyn}
\end{equation}
where: the overdot denotes $d/dt$; $\vec x(t)$ is $m$-dimensional; $\vec 
y(t)$ is $n$-dimensional (more generally, of infinite dimension); the 
eigenvalues of $A$ have zero real-part; the eigenvalues of $B$ have 
strictly negative real-parts bounded away from $0$, 
$\lambda_B<-\gamma\leq0$; and the nonlinear functions $\vec f$ and $\vec 
g$ are smooth and are at least quadratic near the origin.  
Then \cite[p4--5]{Carr81}, \cite[\S2]{Carr83b} or \cite[p5\&p35]{Iooss92}:
\begin{description}
\item[Existence] There exists a smooth $m$-dimensional centre manifold 
for~(\ref{stddyn}) of the form $\vec y=\vec h(\vec x)$, tangent to $\vec 
y=\vec 0$ at the origin.  The dynamics on the centre manifold are governed 
by
\begin{equation}
	\dot{\vec a}=A\vec a+\vec f(\vec a,\vec h(\vec a))\,.
	\label{stdman}
\end{equation}

That is, {\em provided it has a correct spectrum of eigenvalues, a smooth 
dynamical system possesses a centre manifold with 
low-dimensional, self-contained dynamics.}

\item[Relevance] Let $(\vec x(t),\vec y(t))$ be a solution of the parent 
system~(\ref{stddyn}) with initial point $(\vec x(0),\vec y(0))$ 
sufficiently small (in practice, sufficiently small may be quite wide), 
then provided the zero solution is stable, there exists a solution 
of~(\ref{stdman}) $\vec a(t)$ such that as $t\to\infty$,
\begin{equation}
	\vec x(t)=\vec a(t)+\Ord{e^{-\gamma t}}\,,\quad
	\vec y(t)=\vec h(\vec a(t))+\Ord{e^{-\gamma t}}\,,
	\label{stdinf}
\end{equation}
where $\gamma>0$ is some constant.

That is, {\em from a wide range of initial conditions, {\em all} solutions 
tend exponentially quickly to a solution on the centre manifold, and hence 
the dynamical system~(\ref{stdman}), on the $m$-dimensional centre manifold, 
faithfully models the original.}

\item[Approximation] Seeking an approximation to the centre manifold, $\vec 
y=\th(\vec x)$, just substitute into the governing 
equations~(\ref{stddyn}), and use the chain rule to deduce that, given
\begin{equation}
	\cM\th=\D{\th}{\vec x}\left[A\vec x+\vec f\left(\vec x,\th(\vec x)\right)\right]
   -B\th-\vec g\left(\vec x,\th(\vec x)\right)\,,
	\label{stdres}
\end{equation}
we wish to solve $\cM\th=0$.   Suppose that as $\vec x\to\vec 0$, 
$\cM\th=\Ord{|\vec x|^q}$, then $\vec h(\vec x)=\th(\vec x)+\Ord{|\vec 
x|^q}$.  It is often convenient to appeal to a more general assertion 
that explicitly accounts for constant parameters, say $\epsilon$:
if $\cM\th=\Ord{|\vec x|^q,|\epsilon|^r}$, then $\vec 
h(\vec x,\epsilon)=\th(\vec x,\epsilon)+\Ord{|\vec x|^q,|\epsilon|^r}$.  
(This more general form is particularly relevant to unfolding bifurcations and to the 
long-wave, slowly-varying approximation \cite{Roberts88a}.)

That is, {\em provided we can satisfy the governing equations to some order 
of accuracy, then the centre manifold will have been found to the same 
order of accuracy.}
\end{description}
Simple applications of these theorems may be found in the book by 
Carr\cite{Carr81}.  Here I show how to use an algorithm, eminently suitable 
for computer algebra and based upon these theorems, to derive 
low-dimensional models of dynamical systems.  In \S2, I develop the 
formalism in general and in a bifurcation example.  In \S3 I show how 
straightforward it is to apply the techniques to a much more difficult 
problem, namely the flow of a thin film of fluid upon a solid substrate.

One important feature of the analysis is that we deal here with the 
problems in terms of the physical differential equations as given, and {\em 
not} in the abstract form~(\ref{stddyn}).  A linear change of basis, such 
as $(\vec x,\vec y)=P\vec u$ for some linear transformation $P$, is all that 
is needed to transform from a physical description, in terms of physical 
variables $\vec u$, into a form for a direct application of theory.

\section{Modelling a pitchfork bifurcation}

Consider the following variation to Burger's equation featuring growth, 
$(1+\epsilon)u$, nonlinearity, $uu_x$, and dissipation, $u_{xx}$:
\begin{equation}
	\D ut=(1+\epsilon)u+u\D ux+\DD ux\,,
	\quad u(0)=u(\pi)=0\,,
	\label{eburg}
\end{equation}
for some function $u(x,t)$.  View this as an infinite dimensional dynamical 
system, the state space being the set of all functions $u(x)$ on 
$[0,\pi]$.  

For all values of the parameter $\epsilon$ there is a fixed point at the 
origin, that is, a trivial equilibrium state is $u=0$.  A linearisation of 
the equation about this equilibrium, namely $u_t=(1+\epsilon)u+u_{xx}$, has 
modes $\sin kx$ with associated eigenvalues $\lambda_k=1-k^2+\epsilon$ for 
wavenumbers $k=1,2,\ldots$.  Thus the $k=1$ mode, $\sin x$, loses stability 
as $\epsilon$ crosses zero, and the system undergoes a 
bifurcation.

To find the details of this pitchfork bifurcation is a simple task for 
low-dimensional modelling.  Linearly, exactly at critical, $\epsilon=0$, 
all modes decay exponentially quickly except for the critical mode $\sin 
x$; it has a zero decay rate and therefore is long lasting; by the first 
theorem we are assured that there exists a centre manifold.  Nonlinearly, 
and for $\epsilon$ and $u(x)$ near $0$, all modes decay exponentially except for the 
critical modes which have a slow evolution.  Thus, exponentially quickly we 
can model the dynamics solely in terms of the evolution of the amplitude of 
the $\sin x$ mode; I define $a$ to be its amplitude.  By the relevance 
theorem, the evolution of $a$ in time forms an accurate one-dimensional 
model of the original infinite-dimensional dynamical system~(\ref{eburg}).

\subsection{The iteration scheme}

I now proceed to simultaneously develop a novel and powerful algorithm 
to derive such a low-dimensional model while applying it to the specific 
example~(\ref{eburg}).  In general, I address dynamical systems in the form
\begin{equation}
	\dot{\vec u}=\cL{\vec u}+\vec f(\vec u,\epsilon)\,,
	\label{egen}
\end{equation}
where:
\begin{itemize}
\item $\vec u(t)$ is the evolving state ``vector'' (corresponding to 
$(\vec x,\vec y)$ in~(\ref{stddyn})), $u(x,t)$ in the 
example;
	
\item $\cL$ is a linear operator whose spectrum, as required by  
centre manifold theory, is discrete and separates into eigenvalues of zero 
real-part, the critical eigenvalues (corresponding to the modes $\vec x$ 
in~(\ref{stddyn})), and eigenvalues with strictly negative real-part 
(corresponding to the modes $\vec y$ in~(\ref{stddyn})); in the example 
$\cL u=u+u_{xx}$ (with implicit boundary conditions);
	
\item $\epsilon$ is a parameter, potentially a vector of parameters;
	
\item and $\vec f$ is a function which is strictly nonlinear when considered as a 
function in $\vec u$ and $\epsilon$ together, that is, $\vec f$ is 
quadratic or higher order in $\vec u$ and $\epsilon$ as they tend to $0$, 
$\vec f=\epsilon u+uu_x$ in the example.
\end{itemize}
For simplicity, I only treat the case where the critical eigenvalues of 
$\cL$ are exactly $0$; let the multiplicity be $m$.  Cases where the 
critical eigenvalues have a non-zero imaginary part, as in a Hopf 
bifurcation, may be handled with the same ideas as described herein, but 
with more complicating detail.  The aim is to find a low-dimensional model 
$\dot{\vec a}=\vec g(\vec a)$, such as~(\ref{stdman}), for the evolution of 
$m$ ``amplitudes'' $\vec a$.  These low-dimensional dynamics occur on the 
exponentially attractive centre manifold which may be described 
parametrically as $\vec u=\vec v(\vec a)$.

The first stage is to identify the $m$ critical modes, that is, those 
associated with the eigenvalue zero; these are necessary in order to 
project the linear dynamics and nonlinear perturbations onto the slow modes 
of interest.  They may be found from the nontrivial solutions, $\vec v_j$, 
of $\cL\vec v=0$;  in general we need the critical eigenspace and so may 
need to find all the generalised eigen-modes.  
Then, in terms of modal amplitudes $a_j$, a linear approximation to the 
centre manifold and the evolution thereon is simply
\begin{equation}
	\vec u(t)\approx \sum_j\vec v_ja_j=\cV\va
	\quad\mbox{such that}\quad \dot\va\approx \cG\va\,,
	\label{mlin}
\end{equation}
where $\cV=[\vec v_j]$ and where $\cG$ may be chosen in Jordan form in the 
case of generalised eigenvectors (in~(\ref{stddyn}) the $m$ critical modes 
are $\vec x$, and the linear approximation is $\vec u=(\vec x,\vec 0)$ such 
that $\dot{\vec x}=A\vec x$).  In the example~(\ref{eburg}), the eigenvalue 
zero is of multiplicity $m=1$ and so there is only the one critical mode, 
$v(x)=\sin x$; hence the linear approximation is
\begin{equation}
	u(x,t)\approx a\sin x
	\quad\mbox{such that}\quad \dot a\approx 0\,.
	\label{meglin}
\end{equation}
To model the nonlinear dynamics, this linear approximation needs to be 
modified by nonlinear terms; such modification would be equivalent to 
seeking the nonlinear shape of the centre manifold, $\vec y=\vec h(\vec 
x)$, in~(\ref{stddyn}).

The second stage is to seek iterative improvements to a given level of 
description of the centre manifold and the low-dimensional evolution thereon.  The aim is 
to find a low-dimensional description which satisfies the nonlinear 
dynamical equation~(\ref{egen}).  As in Newton's method for finding the 
zero of a function, we use the residual of the governing equations in order 
to guide corrections.  The iteration scheme is successful as long as it 
ultimately drives the residual to zero to the desired order of 
accuracy---see the approximation theorem.  Suppose that at one stage of the 
iteration we have the approximate model
\[ \vec u\approx\tv(\va)
   \quad\mbox{such that}\quad
   \dot\va\approx \tg(\va)\,;
\]
approximate because the residual of the governing differential 
equation~(\ref{egen}) 
\begin{equation} 
\dot{\vec u}-\cL\vec u-\vec f(\vec u,\epsilon)
=\D{\tv}{\vec a}\tg-\cL\tv-\vec f(\tv,\epsilon)
=\Ord{a^q,\epsilon^r}\,,
\label{resid}
\end{equation}
for some order of error, $q$ and $r$, and where $a$ denotes $|\vec a|$.
We seek to find ``small'' corrections, indicated by primes, so that
\[ \vec u\approx\tv(\va)+\pv(\va)
   \quad\mbox{such that}\quad
   \dot\va\approx \tg(\va)+\pg(\va)\,,
\]
is a better approximation to the centre manifold and the evolution thereon.  
The aim of each iteration is to improve the order of the errors ($q$ and 
$r$) so that, by the approximation theorem, we improve the accuracy of the 
model.  Substituting into the governing differential equation~(\ref{egen}), 
and using the chain rule for time derivatives, leads to
\[ \left(\D\tv\va+\D\pv\va\right)\left(\tg+\pg\right)
   =\cL\tv+\cL\pv+\vec f(\tv+\pv,\epsilon)\,.
\]
Given that it is impossible to solve this for the perfect corrections in one 
step, seek an approximate equation for the corrections of 
$\Ord{a^q,\epsilon^r}$ by:
\begin{itemize}
\item ignoring products of corrections (primed quantities) because they 
will be small, $\Ord{a^{2q}+\epsilon^{2r}}$, compared with the dominant 
effect of the linear correction terms;
\item and replacing tilde 
quantities by their initial linear approximation wherever they are 
multiplied by a correction factor (introducing errors 
$\Ord{a^{q+1},\epsilon^{r+1}}$)---such approximation slows the iteration 
convergence to linear, as opposed to the quadratic convergence which would 
be otherwise obtained (if only it were practical).
\end{itemize}
Thus we wish to solve
\[ \D\tv\va\tg +\cV\pg +\D\pv\va\cG\va
   =\cL\tv+\cL\pv+\vec f(\tv,\epsilon)\,.
\]
It is not obvious, but provided it is arranged so that $\cG$ is in Jordan 
form, as is often physically appealing, we may significantly simplify the algorithm by also neglecting the 
term $\D\pv\va\cG\va$ at a cost of increasing the number of iterations needed 
by a factor no more than $m$, the multiplicity of the zero eigenvalue of $\cL$.
Thus, rearranging and recognising that $\D\tv\va\tg=\D\tv t$ by the chain rule, we solve
\begin{equation}
	\cL\pv-\cV\pg=\D\tv t-\cL\tv-\vec f(\tv,\epsilon)\,,
	\label{itgen}
\end{equation}
for the primed correction quantities.  In the example, we seek to solve
\[ v'+\DD{v'}x-g'\sin x=\D{\tilde v}t-(1+\epsilon)\tilde v
   -\tilde v\D{\tilde v}x-\DD{\tilde v}x\,,
\]
which in the first iteration from the linear approximation~(\ref{meglin}) is 
\[ v'+\DD{v'}x-g'\sin x= -\epsilon a\sin x -\frac{1}{2}a^2\sin 2x\,.
\]

The great advantage of this approach is that the right-hand side is 
simply the residual of the governing equation~(\ref{egen}) evaluated at the 
current, tilde, approximation; in essence, this residual is the 
quantity defined in~(\ref{stdres}).  Thus at any iteration we just deal with 
physically meaningful expressions; all the complicated expansions and 
rearrangements of asymptotic expansions, as needed by the method of 
multiple scales (e.g.~\cite[\S3.5]{Jeffrey82}) or earlier methods to find 
the centre manifold (e.g.~\cite[\S5.4]{Coullet83}), are absent.  
There is, of course, a cost and that lies in evaluating the residual (which 
potentially has enormous algebraic detail), much of which is repeated at every 
iteration.  However, with the advent of computer algebra, all this detail 
may be left to the computer to perform---such mindless repetition is ideal 
for a computer---whereas all a human need concern themselves with is 
setting up the solution of~(\ref{itgen}) and not at all with the detailed 
algebraic machinations of asymptotic expansions.

The main detail is to solve equations of the form
\begin{equation}
	\cL\pv-\cV\pg=\vec r\,,
	\label{iteqn}
\end{equation}
for some given residual $\vec r$.  Recognise that there are more unknowns 
than components in this equation; its solution is not unique.  The freedom 
comes from the fact that we can parameterise the centre manifold via the 
amplitudes $\vec a$ in an almost arbitrary manner.  The freedom can only be 
resolved by giving a precise meaning to the $m$ amplitudes $\vec a$.  Often 
one does define $\vec a$ to be precisely the modal amplitudes (as is done 
implicitly for~(\ref{stddyn}) by seeking a centre manifold in the form 
$(\vec x,\vec h(\vec x))$) in which case 
we seek corrections $\pv$ which are orthogonal to the generalised 
eigenvalues, $\vec z_j$, of the adjoint of $\cL$.  In the example, $\cL$ is 
self adjoint under the inner product $\langle u_1,u_2\rangle =\frac{2}{\pi}
\int_0^\pi u_1u_2dx$, and so the adjoint eigenvector is also simply $z(x)=\sin 
x$ and so I require $\langle \sin x,v'\rangle=0$.  More general definitions, 
such as an energy related amplitude, give rise to similar considerations to 
those that follow.  There are two approaches to solving~(\ref{iteqn}).
\begin{enumerate}
\item Numerically, it is generally easiest to adjoin the amplitude condition to 
the equation and solve 
\[ \left[
\begin{array}{cc}
	\cL & -\cV  \\
	\cZ^T & 0
\end{array}
\right]\left[
\begin{array}{c}
	\pv  \\
	\pg
\end{array}
\right] = \left[
\begin{array}{c}
	\vec r  \\
	\vec 0
\end{array}
\right]\,,
\]
where $\cZ=\left[\vec z_j\right]$.
	
\item However, algebraically it is usually more convenient to adopt the 
following procedure which is also familiar as part of other asymptotic methods.  
Rewrite~(\ref{iteqn}) as $\cL\pv=\cV\pg+\vec r$ and recognise that 
$\cL$ is singular due to the zero eigenvalue of multiplicity $m$.  We 
choose the $m$ components of $\pg$ to place the right-hand side in the 
range of $\cL$; this is achieved by taking the inner product of the 
equation with the adjoint eigenvalues $\vec z_j$ (this corresponds to 
considering just the $\vec x$ modes in~(\ref{stddyn})) and thus giving the set 
of $m$ equations 
\[  \langle\cZ,\cV\rangle\pg=-\langle\cZ,\vec r\rangle\,.
\]  
In the example, $g'=-\frac{2}{\pi}\int_0^\pi\sin x\,r(x)\,dx$ which in 
the first iteration gives $g'=\epsilon a$.  This equation is 
known as the solvability condition.

Having put the right-hand side in the range of $\cL$ we solve 
$\cL\pv=\hat{\vec r}=\cV\pg+\vec r$ for $\pv$, making the solution unique 
by accounting for the definition of the amplitudes $\va$.  In the 
example, we solve the boundary value problem $v'+{v'}_{xx}=\hat r(x)$ 
such that $v'(0)=v'(\pi)=0$ and that $v$ has no $\sin x$ component.  In the 
first iteration, the problem $v'+{v'}_{xx}=-\frac{1}{2}a^2\sin 2x$ with 
the above conditions has the solution $v'=\frac{1}{6}a^2\sin 2x$.
\end{enumerate}
Then the last step of each iteration is to update the approximations for 
the centre manifold shape and the evolution thereon.  For the example, 
after the first iteration we deduce $u\approx a\sin x+\frac{1}{6}a^2\sin 
2x$, which shows the nonlinear steepening/flattening of negative/positive 
slopes, and that the evolution is $\dot a\approx \epsilon a$, which 
exhibits the loss of stability of the fixed point $a=0$ as $\epsilon$ 
becomes positive.

Further iterations in the example lead to the centre manifold being given by
\begin{equation}
	u=a\sin x+\frac{1-\frac{\epsilon}{3}}{6}a^2\sin 2x
	+\frac{1-\frac{\epsilon}{12}}{32}a^3\sin 3x
	+\Ord{a^4,\epsilon^2}\,,
	\label{burgu}
\end{equation}
on which the system evolves according to
\begin{equation}
	\dot a=\epsilon a-\frac{1-\frac{\epsilon}{3}}{12}a^3+\Ord{a^4,\epsilon^2}\,.
	\label{burgamp}
\end{equation}
The relevance theorem assures us that this 1-D model of the original 
infinite-dimensional dynamical system~(\ref{eburg}), is valid exponentially 
quickly in time.  From the model~(\ref{burgamp}), for example, we deduce 
the quantitative shape of the pitchfork bifurcation: there are stable fixed 
points at $a\approx\sqrt{\epsilon/(1-\epsilon/3)}$.  Physically, these 
fixed points represent a balance between the nonlinear steepening of the 
$uu_x$ term, and the dissipation of $u_{xx}$.

\subsection{Computer algebra implementation}

A principal reason for adopting this approach is because it is simply and 
reliably implemented in computer algebra.  Based upon the above 
derivation, the general outline of the algorithm is:
\begin{enumerate}\sf
	\item  preliminaries;
	
	\item  initial linear approximation;
	
	\item  {\tt repeat until} residual is small enough;
	\begin{enumerate}
		\item  compute residual,
		
		\item  find solvability condition,
		
		\item  compute correction to the centre manifold,
		
		\item  update approximations.
	\end{enumerate}
\end{enumerate}
Complete details of a {\reduce} program for the particular 
example~(\ref{eburg}) follows.  The reason for using {\reduce}\footnote{At 
the time of writing, information about {\reduce} was available from Anthony 
C.\ Hearn, RAND, Santa Monica, CA 90407-2138, USA.  E-mail: \tt 
reduce@rand.org} is that it has excellent pattern matching and replacement 
capabilities through its {\tt operator} and {\tt let} statements.
\begin{verbatim}
% Find pitchfork bifurcation in u_t=(1+eps)u+uu_x+u_xx
% a(t) measures amplitude of sin(x) component in u(x,t)
%
on div; off allfac; on revpri;  % improves appearance of output
let sin(~x)*cos(~y) => (sin(x+y)+sin(x-y))/2; % a trig rule 
%
% Define the inverse operator of  u+u_xx
operator linv; linear linv;
let linv(sin(~k*x),x) => sin(k*x)/(1-k^2);
% Define inner product with sin(x)
operator sindot; linear sindot;
let { sindot(sin(x),x) => 1, sindot(sin(~k*x),x) => 0 };
% 
depend a,t;         % asserts that  a  depends upon time t
let df(a,t) => g;   % so da/dt is replaced by current g(a,eps)
%
u:=a*sin(x); g:=0;  % initial approximation
%
% iterate until PDE is satisfied (to requisite order)
let {eps^2=0, a^4=0};  % discard high-order terms in a & eps
repeat begin
    write eqn:=df(u,t)-(1+eps)*u-u*df(u,x)-df(u,x,x);
    gd:=-sindot(eqn,x);
    write u:=u+linv(eqn+sin(x)*gd,x);
    write g:=g+gd;
end until eqn=0;
%
;end;
\end{verbatim}
Observe the how this implements the algorithm.
\begin{enumerate}
\item  The preliminaries do the following.
    \begin{itemize}
    
\item $\ell4$ improves the appearance of the printed output, whereas 
$\ell$5 tells {\reduce} that we wish to linearise products of trigonometric 
functions.
		
		\item  $\ell$7--9 defines the operator {\tt linv} to act as the inverse 
		of $\cL$:
        \begin{itemize}
			\item  declaring it {\tt linear} tells {\reduce} to expand sums and 
			products in the first argument and to only leave functions of the 
			second argument inside the operator, for example, {\tt linv($\epsilon 
			a\sin x+2a^2\sin 2x$,x)} is expanded to
			\\ {\tt $\epsilon a$linv($\sin 
			x$,x)+$2a^2$linv($\sin 2x$,x)};
			
			\item  the {\tt let} statement on $\ell$9 defines the action of the operator as 
			the solution to $v'+v'_{xx}=\sin kx$, namely $v'=\frac{1}{1-k^2}\sin kx$, 
			the tilde before the {\tt k} on the left-hand side matches any pattern
			(no action is defined for the singular case $k=1$ because the 
			pattern {\tt sin(\verb|~|k*x)} does not match {\tt sin(x)}, but 
			any appearance of {\tt linv(sin(x),x)} usefully signals an error). 
		\end{itemize}
		
		\item  $\ell$10--12 similarly defines {\tt sindot} to be the inner 
		product operator $\langle\sin x,\cdot\rangle$, the {\tt let} 
		statement now being a list, enclosed within braces, of evaluation 
		rules. 
		
		\item $\ell$14--15 establishes that the variable {\tt a} is to firstly 
		depend upon time, as we use $a$ as the time dependent amplitude in the 
		model, and secondly that time derivatives of $a$, {\tt df(a,t)}, are to 
		be replaced by the value of {\tt g}, at the time of replacement, as 
		{\tt g} is to store the current approximate model evolution equation such 
		as~(\ref{burgamp}).
	\end{itemize}
	
	\item  $\ell$17 simply assigns the linear approximation~(\ref{meglin}) of the 
	centre manifold to be the initial value of the variables {\tt u} and {\tt g}.
	
	\item  $\ell$20--26 performs the iterations.
	\begin{itemize}
		\item  $\ell$20 controls the truncation of the asymptotic 
		approximation.  It gives a list of transformations which tell
		{\reduce} to discard any factor in $\epsilon^2$ or higher and any factor 
		in $a^4$ or higher;  thus all expressions are computed to an error of
		$\Ord{\epsilon^2,a^4}$. 
		
		\item  $\ell$22 computes the residual; observe how it is a very direct 
		translation of the governing equation~(\ref{eburg}) into {\reduce} 
		symbols, a user of this approach only has to implement the governing 
		equations, all the messy details of the asymptotic expansions are  
	    dealt with by the computer algebra engine. 
		The iteration is repeated until this residual is zero, 
		to $\Ord{\epsilon^2,a^4}$, $\ell$26.
		
		\item  $\ell$23 is the solvability condition giving corrections to {\tt g}.
		
		\item  $\ell$24--25 solves for the correction to {\tt u} and updates 
		the current approximation.
	\end{itemize}
\end{enumerate}
The program could be used to derive high-order effects in $a$ or 
$\epsilon$ simply by changing the discarding of factors on $\ell$20.  
Different systems with the same linear structure may be analysed simply by 
changing the computation of the residual on $\ell$22.

By way of comparison, Rand \& Armbruster's \cite[pp27--34]{Rand87} {\sc 
macsyma} code for constructing the centre manifold of a finite-dimensional 
dynamical system, based upon power series manipulation, uses 53~lines of 
active code (although some are for input of the dynamical equations) whereas the 
above algorithm has only 16~lines of active code.

\section{Thin film fluid dynamics}

We now turn to the modelling of an important physical problem, that of the 
flow of a thin film of viscous fluid upon a solid substrate.  Examples 
include the flow of rainwater on a road or windscreen or other draining 
problems \cite{Chang94}, paint and coating flows \cite{Ruschak85,Tuck90a}, 
and the flow of many protective biological fluids \cite{Grotberg94}.  For 
simplicity we restrict attention to 2D fluid flow and seek a model for the 
evolution of the depth of the fluid film; because the film is thin expect 
that the vertical structure of the flow is relatively simple.  I show how 
centre manifold theory can partially justify such a model, and how to 
derive the model using the algorithm developed in this paper.

\begin{figure}
	\begin{center}
{\tt    \setlength{\unitlength}{0.92pt}
\begin{picture}(370,141)
\thinlines    \put(20,19){\vector(0,1){107}}
              \put(13,28){\vector(1,0){63}}
\thicklines   \put(82,28){\line(1,0){278}}
\thinlines    \put(73,14){$x$}
              \put(10,122){$y$}
              \put(159,12){solid, $u=v=0$}
              \put(24,88){\line(1,0){34}}
              \put(58,88){\line(6,1){47}}
              \put(105,96){\line(3,1){38}}
              \put(143,109){\line(6,1){43}}
              \put(186,115){\line(1,0){35}}
              \put(221,115){\line(6,-1){48}}
              \put(269,107){\line(1,0){38}}
              \put(307,107){\line(6,-1){49}}
              \put(49,99){$y=\eta(x,t)$}
              \put(138,119){atmospheric pressure}
              \put(170,77){Navier-Stokes equation}
              \put(341,59){$h$}
              \put(205,56){$p(x,y,t)$}
              \put(115,56){$\vel(x,y,t)$}
              \put(337,64){\vector(0,1){33}}
              \put(337,65){\vector(0,-1){36}}
              \put(89,88){\vector(1,0){25}}
              \put(89,75){\vector(1,0){22}}
              \put(89,62){\vector(1,0){19}}
              \put(89,49){\vector(1,0){13}}
              \put(89,36){\vector(1,0){6}}
\end{picture}}
	\end{center}
	\caption{schematic diagram of a thin fluid film flowing down a solid bed.}
	\protect\label{svffilm}
\end{figure}
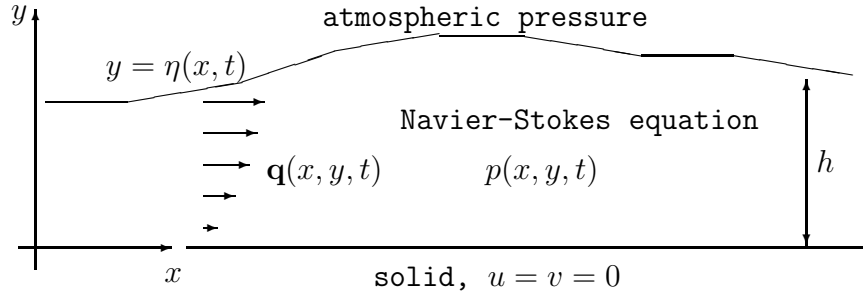

This fluid flow problem has many nonlinearities, as shown in 
Figure~\ref{svffilm}: not only is the advection in the Navier-Stokes 
equation described by a nonlinear term, but also the thickness of the fluid 
film is to be found as part of the solution and its unknown location is 
another source of nonlinearity.  When a fluid layer is thin then, as in 
dispersion in a pipe \cite{Mercer94a}, the dominant dynamical processes 
occur along the thin film of fluid.  Across the fluid film, viscosity acts 
quickly to damp almost all cross-film structure; the only long lasting 
dynamics are those of the relatively slow spread of the film along the 
substrate.  In a nonlinear problem, this signature of interesting dynamics 
on a long time-scale along with uninteresting, quickly dissipated modes 
indicates that centre manifold theory may be used to create a 
low-dimensional model of the interesting dynamics.

Assuming no longitudinal variations in $x$, a linear analysis of the 
equations shows that there is one critical mode in the cross-film dynamics 
associated with conservation of the fluid, all others decay due to 
viscosity.  Consequently, it is natural to express the low-dimensional 
model in terms of the film thickness $\eta(x,t)$.  Seeking solutions which 
vary slowly along the film, that is, assuming $\partial/\partial x$ is a 
small parameter like $\epsilon$ used previously, a centre manifold analysis 
creates an effective model of the dynamics.  Through the algorithm 
described herein, we find the long-term evolution of long-wavelength modes 
is approximately described by
\begin{eqnarray}
	\Dt\eta & \approx & -\frac{1}{3}\Dx{\ }\left(\eta^3\eta'''\right)  
	\label{svenm}\\
	 &  & -\Dx{\ }\left[\frac{3}{5}\eta^5\eta^v +3\eta^4\eta'\eta^{iv} 
	 +\eta^4\eta''\eta''' +\frac{11}{6}\eta^3{\eta'}^2\eta''' 
	 -\eta^3\eta'{\eta''}^2 \right]\,,
	 \nonumber
\end{eqnarray}
where dashes or roman numerals on $\eta$ denote $\partial/\partial x$.
The centre manifold theorems reasonably assure us that this model is indeed 
relevant to the long-term dynamics of thin films.  However, this assurance 
is not yet rigorous because of deficiencies in the preconditions of current 
theorems.\footnote{The major constraint upon a rigorous application of 
the theorems by Gallay \cite{Gallay93} is that we need to treat 
$\partial/\partial x$ as a small parameter, whereas in principle it is 
unbounded.  However, provided we confine attention to functions 
slowly-varying in $x$, then such derivatives would indeed be small.}

The first line of the model~(\ref{svenm}) is the standard leading, or 
``lubrication,'' approximation to thin film dynamics; the second line may 
be viewed either as correction terms, or as terms indicative of the error 
in the leading approximation.  An interesting diversion is attributable to 
the fact that although this is a nonlinear problem, conservation of fluid 
applies no matter how thick the fluid layer (for example, the right-hand 
side of~(\ref{svenm}) can always be written as a gradient).  Thus the 
analysis is valid for arbitrarily large variations in the thickness of the 
film! just so long as the variations are sufficiently slow.  Theorems on 
the existence and approximation of such global (in $\eta$) centre manifolds 
are also given by Carr \cite[pp31--32]{Carr81}.

\subsection{Governing equations}

As shown in Figure~\ref{svffilm} we solve the Navier-Stokes and continuity 
equations within the Newtonian fluid for the velocity field $\vel$:
\begin{eqnarray*}
	\Dt\vel+\vel\cdot\grad\vel 
	& = & -\frac{1}{\rho}\grad p+\nu\grad^2\vel\,,  \\
	\divv\vel & = & 0\,.
\end{eqnarray*}
For simplicity, restrict attention to two-dimensional flow taking place 
in the $xy$-plane: the $x$-axis is aligned along the solid bed of 
the flow; the $y$-axis is perpendicular.  The viscous flow must stick to 
the solid bed to give the boundary condition
\begin{equation}
	\vel=\vec 0\quad\mbox{on $y=0$.}
	\label{svebbc}
\end{equation}

The surface of the fluid, $y=\eta(x,t)$, evolves with the flow.  Because 
the free-surface is unknown we not only need two boundary conditions for 
the Navier-Stokes equations, we also need an extra boundary equation in 
order to be able to find $\eta$.  The kinematic condition is that the fluid 
flow, as given by the velocity, $\vel$, must follow the free-surface as 
given by $y=\eta(x,t)$:
\begin{equation}
     \Dt\eta = v-u\Dx\eta\quad\mbox{on $y=\eta$.}
	\label{svekc}
\end{equation}
The other boundary conditions come from the forces acting across the 
free-surface.  Above the fluid film we suppose there is a very light 
and essentially inviscid fluid such as air (one-thousandth the density of 
water).   
\begin{itemize}
\item Since air is inviscid, it cannot sustain any tangential stress across 
the surface, thus
\begin{equation}
	2\eta'\left(\Dy v-\Dx u\right) 
+\left(1-\eta'^2\right)\left(\Dy u+\Dx v\right) =0
\quad\mbox{on $y=\eta$.}
	\label{svetc}
\end{equation}
	
\item Since the density of air is very low, Bernoulli's equation asserts 
that fluid stress exerted normally across the surface has to be constant, 
say $T_n=-p_a$, equal and opposite to air pressure which without loss of 
generality I take to be zero.  However, the effect of surface tension is 
like that of an elastic membrane, it causes a pressure jump if the surface 
is curved: positive if the fluid surface is convex; negative if it is 
concave.  For both the normal stress and surface tension to oppose atmospheric 
pressure
\begin{eqnarray*}
	\left(1+\eta'^2\right)p & = & 2\mu\left[\Dy v+\eta'^2\Dx 
	u-\eta'\left(\Dy u+\Dx v\right)\right]   \\
	 &  & -\frac{\sigma\eta''}{\sqrt{1+\eta'^2}} 
\quad\mbox{on $y=\eta$,}
\end{eqnarray*}
where $\sigma$ is the coefficient of surface tension.
\end{itemize}

Now non-dimensionalise these equations, refering all scales to $\sigma$ 
as surface tension is taken to be the driving force.  For a reference length, suppose 
that $h$ is a characteristic thickness of the thin film as shown 
schematically in Figure~\ref{svffilm}.  Then non-dimensionalise by writing 
the equations with respect to: the reference length $h$; the reference time 
$\mu h/\sigma$; the reference velocity 
$U=\sigma/\mu$, and the reference pressure $\sigma/h$.
With these choices, and in non-dimensional quantities, we solve the 
Navier-Stokes and continuity equations
\begin{eqnarray}
	\cR\left(\Dt \vel+\vel\cdot\grad \vel\right) & = & -\grad p+\grad^2\vel \,,
	\label{svefq} \\
	\divv\vel & = & 0\,,
	\label{svefc}
\end{eqnarray}
where $\cR=\frac{\sigma h}{\mu\nu}$ is effectively a Reynolds number 
characterising the importance of the inertial terms---it may be written 
as $Uh/\nu$ for the above characteristic velocity.  Observe that due to the absence of $\cR$ in the 
model~(\ref{svenm}), to its order of accuracy the inertial terms have no 
influence on the dynamics---it is creeping flow.  The non-dimensional 
equations are subject to the bottom boundary condition~(\ref{svebbc}), and 
on the surface the kinematic condition~(\ref{svekc}) and the two dynamic 
conditions, (\ref{svetc}) and
\begin{equation}
	\left(1+\eta'^2\right)p=
	2\left[\Dy v+\eta'^2\Dx u-\eta'\left(\Dy u+\Dx v\right)\right]
  -\frac{\eta''}{\sqrt{1+\eta'^2}} 
\quad\mbox{on $y=\eta$.}
	\label{svenc}
\end{equation}

In the Navier-Stokes equation I have neglected body forces.  If the 
presence of gravity were to be acknowledged, then it would appear in the 
non-dimensional combination of the Bond number $b=g\rho h^2/\sigma$.  This 
may be neglected if $b$ is extremely small ($h\ll 0.2\cm$ for water).  
Including gravity, assuming $b$ is small but not negligible, is not a 
significant complication and I leave its inclusion as an exercise for the 
reader---perhaps the simplest interesting case is that of a fluid film on a 
vertical substrate.

\subsection{Linear picture}

The construction of a centre manifold model rests upon an understanding of the 
dynamics linearised about some fixed point (the spectrum of the linear 
operator is crucial in theory and application).  Here the fixed point is a film 
of constant thickness, non-dimensionally $1$, and of zero velocity and 
pressure.  The dynamics linearised about this fixed point are based on 
velocities and $\eta-1$ being small.  Thus, as well as neglecting products 
of small terms, boundary conditions on the free-surface are 
approximated by their evaluation on the approximate surface $y=1$.  The 
linearised equations are then
\begin{eqnarray}
	\cR\Dt\vel & = & -\grad p+\grad^2\vel\,,  \nonumber\\
	\divv\vel & = & 0\,,  \nonumber\\
	\vel & = & \vec 0\quad\mbox{on $y=0$,} \label{svelf}\\
	\Dt\eta & = & v\quad\mbox{on $y=1$,} \nonumber\\
	\Dy u+\Dx v & = & 0\quad\mbox{on $y=1$,} \nonumber\\
	p & = & 2\Dy v-\eta''\quad\mbox{on $y=1$.} \nonumber
\end{eqnarray}

To investigate the dynamics of ``long waves'' on this thin film we treat 
$\partial/\partial x$ as small, that is slowly-varying in $x$, as justified 
formally in \cite{Roberts88a} and more rigorously for capillary-gravity 
waves in \cite{Haragus95}.  ``Linearly'' then we can base the construction 
of a centre manifold model on the limit when there are {\em no} 
longitudinal variations: $\Dx{\ }=0$.  Neglecting all $x$ derivatives and 
seeking solutions proportional to $e^{\lambda t}$ leads to the following.  
Firstly, $\lambda_0=0$ is an eigenvalue associated with the mode 
$\eta=\mbox{const.}$ and $u=v=p=0$.  It is the presence of this eigenvalue 
which indicates the existence of a centre manifold model.  Other modes 
exist, namely $u_n=\sin\pi(n-\half)y$, $v_n=p_n=\eta_n-1=0$ with 
eigenvalues $\lambda_n=-\pi^2(n-\half)^2/\cR$ for $n=1,2,\ldots$.  Thus for 
long-waves there exists one $0$ eigenvalue of the dynamics, whereas all the 
rest of the eigenvalues are strictly negative, bounded above by 
$-\gamma\approx -2.5/\cR$ (the smaller the Reynolds number, the more 
creeping the flow, and the faster the decay of the non-critical modes).  
Hence, by the relevance theorem, expect the dynamics of thin films to 
exponentially quickly approach a low-dimensional centre manifold based on 
the mode corresponding to the $0$ eigenvalue, the film thickness $\eta$, 
and characterised by slow variations in $x$.

\subsection{Iterative construction}

The first task is to decide how to parameterise the centre manifold model.  
Since the critical mode is $\eta=\mbox{const.}$, $p=u=v=0$, it is 
appropriate to use the film thickness $\eta$ as the parameter.  This is 
especially useful since $\eta$ has a direct physical meaning.  Thus a 
linear description of the centre manifold is
\begin{equation}
	u=v=p=0\,,
	\label{svefln}
\end{equation}
and $\eta$ free to vary according to $\Dt\eta\approx 0$.

Now we organise an iteration scheme to refine the description of the centre 
manifold.  The order of error will be characterised by the number of 
spatial derivatives of $\eta$; for a slowly-varying function higher-order 
derivatives are asymptotically smaller than lower-order derivatives.  
Suppose that the fields $\tilde\vel(\eta,y)$ and $\tilde p(\eta,y)$ are an approximation 
to the centre manifold ``shape,'' with the associated approximate evolution 
$\Dt\eta\approx\tilde g(\eta)$.  Seek equations for corrections to this 
description by:
\begin{itemize}
\item substituting into the governing equations:\footnote{Be careful not to 
confuse corrections, indicated by primes, and derivatives of $\eta$ with 
respect to $x$, also indicated by primes.}
\begin{eqnarray*}
	\vel & = & \tilde\vel(\eta)+\vel'(\eta)\,, \\
	p & = & \tilde p(\eta)+p'(\eta)\,,  \\
	\mbox{such that}\quad \Dt\eta & = & \tilde g(\eta)+g'(\eta)\,;
\end{eqnarray*}
	
\item omit products of corrections;

\item omit $x$-derivatives of corrections as both corrections and 
``$\Dx{\ }$'' are small;
	
\item approximate other terms involving corrections by replacing the 
current approximation, tilde quantities, with the initial linear 
approximation (here zero);
	
\item rearrange the equations.
\end{itemize}
The upshot is that we solve equations
\begin{eqnarray}
	\Dyy{u'} & = & \cR\left(\Dt{\tilde u}+\tilde\vel\cdot\grad \tilde u\right) 
	+\Dx{\tilde p} -\grad^2\tilde u\,,
	\label{sveisu} \\
	\Dyy{v'}-\Dy{p'} & = & \cR\left(\Dt{\tilde v}+\tilde\vel\cdot\grad \tilde v\right) 
	+\Dy{\tilde p} -\grad^2\tilde v\,,
	\label{sveisv} \\
	\Dy{v'} & = & -\divv{\tilde\vel}\,,
	\label{sveisc} 
\end{eqnarray}
with boundary conditions
\begin{eqnarray}
	\Dy{u'} & = & 
	-\left(1-\eta'^2\right)\left(\Dy {\tilde u}+\Dx {\tilde v}\right)
	-2\eta'\left(\Dy {\tilde v}-\Dx {\tilde u}\right) 
    \quad\mbox{on $y=\eta$}\,,
	\label{sveistt} \\
	p'-2\Dy{v'} & = &	-\left(1+\eta'^2\right){\tilde p}
	+2\left[\Dy {\tilde v}+\eta'^2\Dx {\tilde u}
	       -\eta'\left(\Dy {\tilde u}+\Dx {\tilde v}\right)\right]
	       \nonumber\\&&\quad
    -\frac{\eta''}{\sqrt{1+\eta'^2}} 
	\quad\mbox{on $y=\eta$}\,,
	\label{sveistn} \\
	\vel' & = & \vec 0
	\quad\mbox{on $y=0$}\,,
	\label{sveis0} \\
	\tilde g+g' & = & \tilde v+v'-(\tilde u+u')\Dx\eta
	\quad\mbox{on $y=\eta$}\,.
	\label{sveish}
\end{eqnarray}

However, the unknown location of the free surface of the film is a major 
technical difficulty.  One way to proceed is to scale the vertical 
coordinate, $\zeta=y/\eta$, so that the free surface corresponds to 
$\zeta=1$ precisely.  Because $\eta$ varies with $x$ and 
$t$, this scaling of $y$ affects space-time derivatives and so plays havoc 
with details of the governing equations.  Under 
the change of coordinates from $(x,y,t)$ to
\[ \xi=x\,,\quad \zeta=y/\eta(x,t)\,,\quad \tau=t\,,
\]
the chain rule shows that derivatives transform according to
\begin{equation}
	\Dx{\ } = \D{\ }\xi-\zeta\frac{\eta'}{\eta}\D{\ }\zeta\,,  \quad
	\Dt{\ } = \D{\ }\tau-\zeta\frac{\dot\eta}{\eta}\D{\ }\zeta\,,  \quad
	\Dy{\ } = \frac{1}{\eta}\D{\ }\zeta\,.
\label{svectr}
\end{equation}
Fortunately we may implement the analysis in computer algebra and thus 
relegate virtually all such details to the computer.

The only places where we need to explicitly consider these rules are in 
the terms in the equations for the corrections, $u'$, $v'$, and $p'$.  
However, these only involve $y$ derivatives, 
see~(\ref{sveisu}--\ref{sveistn}), which simply transform $\D{\ 
}y\to\frac{1}{\eta}\D{\ }\zeta$.  Thus, we multiply the residuals on the 
right-hand sides of these equations by the appropriate power of $\eta$, as 
seen in lines~40, 47 and 53 of the following {\reduce} program.
\begin{verbatim}
% Construct slowly-varying centre manifold of thin film fluids.
% Allows for large changes in film thickness.
%
on div; off allfac; factor d,h,re; % improves printing 
% use stretched coordinates: ys=y/h(x,t), xs=x, ts=t
depend xs,x,y,t;
depend ys,x,y,t;
depend ts,x,y,t;
let { df(~a,x) => df(a,xs)-ys*h(1)/h(0)*df(a,ys)
    , df(~a,t) => df(a,ts)-ys*g/h(0)*df(a,ys)
    , df(~a,y) => df(a,ys)/h(0) 
    , df(~a,x,2) => df(df(a,x),x) };
% solves -df(p,ys)=rhs s.t. sub(ys=1,p)=0
operator psolv; linear psolv;
let {psolv(ys^~n,ys) => (1-ys^(n+1))/(n+1)
    ,psolv(ys,ys) => (1-ys^2)/2
    ,psolv(1,ys) => (1-ys) };
% solves df(u,ys,2)=rhs s.t. sub(ys=0,u)=0 & sub(ys=1,df(u,y))=0
operator usolv; linear usolv;
let {usolv(ys^~n,ys) => (ys^(n+2)/(n+2)-ys)/(n+1)
    ,usolv(ys,ys) => (ys^3/3-ys)/2
    ,usolv(1,ys) => (ys^2/2-ys) };
% use operator h(m) to denote df(h,x,m)
operator h;
depend h,xs,ts;
let { df(h(~m),xs) => h(m+1)
     ,df(h(~m),xs,2) => h(m+2)
     ,df(h(~m),ts) => df(g,xs,m) };
%
% linear solution
u:=0; v:=0; p:=0; g:=0;
%
% Iteration.  Use d to count the number of derivatives of x,
% and throw away this order or higher in d/dx
let d^7=0;
curv:=h(2)*d^2*(1-h(1)^2*d^2/2+3*h(1)^4*d^4/8-5*h(1)^6*d^6/16);
repeat begin
   % continuity & bed
   write ceq:=-df(u,x)*d-df(v,y);
   write v:=v+h(0)*int(ceq,ys);
   % vertical momentum & normal stress
   write veq:=re*( df(v,t)+u*df(v,x)*d+v*df(v,y) )
              +df(p,y) -df(v,x,2)*d^2-df(v,y,2);
   write tn:= sub(ys=1,-p*(1+h(1)^2*d^2) +2*(df(v,y)
                 +h(1)^2*d^3*df(u,x)-h(1)*d*(df(u,y)+df(v,x)*d))
                 -curv );
   write p:=p+h(0)*psolv(veq,ys) +tn;
   % horizontal momentum & bed & tangential stress
   write ueq:=re*( df(u,t)+u*df(u,x)*d+v*df(u,y) )
              +df(p,x)*d -df(u,x,2)*d^2-df(u,y,2);
   write tt:=-sub(ys=1, (1-d^2*h(1)^2)*(df(u,y)+df(v,x)*d)
                 +2*h(1)*d*(df(v,y)-df(u,x)*d) );
   write u:=u+h(0)^2*usolv(ueq,ys)+h(0)*tt*ys;
   write g:=sub(ys=1,v-u*h(1)*d);
end until (veq=0)and(tn=0)and(ueq=0)and(tt=0)and(ceq=0);
;end;
\end{verbatim}

Recognise the algorithm in this code.  
\begin{enumerate}
	\item  {\sf Preliminaries}
	\begin{itemize}
		\item  The change of coordinate rules~(\ref{svectr}) are implemented in 
$\ell$5--12, where {\tt xs} denotes $\xi$, {\tt ys} denotes $\zeta$, and 
{\tt ts} denotes $\tau$.  Then wherever in the algebraic details we need 
these rules they are automatically invoked by {\reduce}.  For example, in 
the evaluation of the residuals of the nonlinear equations and in boundary 
conditions within the iteration.

\item $\ell$13--17 defines the operator {\tt psolv} which is used to find 
the pressure correction through~(\ref{sveisv}).  For each term in $\zeta^n$ 
in the right-hand side, it integrates with the correct integration constant 
to give a contribution to the pressure.
		
\item $\ell$18--22 defines the operator {\tt usolv} which is used to find 
the horizontal velocity correction through~(\ref{sveisu}).  For each term 
of the form $\zeta^n$, it solves an {\sc ode} with the appropriate bottom 
and linearised free-surface boundary condition.
		
\item $\ell$23--28 For compactness of the output, it is convenient to 
represent the film thickness and its derivatives via an operator {\tt h}, 
$\ell$24.  {\tt h} is to depend upon $x$ and $t$, $\ell$25, and {\tt h(m)} 
denotes $\partial^m\eta/\partial x^m$, for example {\tt h(0)} and {\tt 
h(1)} denote $\eta$ and $\eta'$ respectively, and so $x$ derivatives 
operate according to the transformations on $\ell$26--27.  Whereas 
$\ell$28 encodes the fact that 
\[ \D{\ }t\left(\Dn \eta x m\right)=\Dn{\ }x m\left(\Dt\eta\right)
=\Dn g x m\,,
\]
where the dependence of $\eta$ upon $x$ is recognised in the spatial 
derivatives of $g$.
\end{itemize} 
Note that there is no need to define a specific operator to compute the 
correction to the vertical velocity, from~(\ref{sveisc}), because it is 
obtained simply by integrating the right-hand side as implemented by the 
standard {\tt int} operator in {\reduce}.
	
	\item  {\sf The linear approximation} is specified in $\ell$31.
	
\item {\sf Iteration.} The iteration is carried out until the errors are 
$\Ord{\partial^7/\partial x^7}$, $\ell$35, whether in the form $\eta^{iiv}$ 
or as $\eta''\eta^v$ or as some other such nonlinear combination of 
derivatives of the film thickness $\eta$.  This is achieved by carrying a 
dummy variable {\tt d} whose exponent counts the number of $x$ 
derivatives, and discarding any term of order 7 or more in {\tt d}.

$\ell$36 establishes the expansion for 
$\eta''/\sqrt{1+\eta'^2}$ in a series in small slope 
$\eta'$.  Note the use of {\tt d} to count the total number of $x$ 
derivatives in each term.
	
One variation in the iteration is that the loop, $\ell$37--55, uses new 
information as it becomes available: 
\begin{enumerate}
\item first, the $v$ correction from continuity, $\ell$38--40;
	
\item second, the pressure correction is found, $\ell$40--47, from the 
vertical momentum residual and the normal stress across the 
free-surface (the term in $v'$ on the left-hand side of~(\ref{sveisv}) is 
included in the right-hand side of $\ell$43 because we use the latest 
approximation to $v$);
	
\item third, the $u$ correction is found, $\ell$48--53, from the 
horizontal momentum residual and the tangential stress across the 
free-surface;
	
\item and lastly, the latest version of the model evolution $\Dt\eta=g$ 
from the kinematic free surface condition, $\ell$54.
\end{enumerate}
\end{enumerate}

Upon executing this program, I find not only the evolution 
equation~(\ref{svenm}), but also the velocity and pressure fields.   
Specifically, to errors of fifth-order in $\partial/\partial x$,
\begin{eqnarray*}
	u & \approx & \left(\zeta-\frac{1}{2}\zeta^2\right)\eta^2\eta'''\,,  \\
	v & \approx & -\frac{1}{2}\zeta^2\eta^2\eta'\eta''\,,  \\
	p & \approx & -\eta''  \\
	 &  & +\frac{3}{2}{\eta'}^2\eta''-\left(1+\zeta\right)\eta\eta'\eta''' 
	 -\left(\frac{1}{2}+\zeta-\frac{1}{2}\zeta^2\right)\eta^2\eta^{iv}\,. 
\end{eqnarray*}
Expressions such as these inform us of the dominant details of the flow 
modelled by~(\ref{svenm}); higher-order terms were computed but I have not 
recorded them here.

\subsection{Aside}

Although this model describes the long-term dynamics of thin films, it is 
limited in its usefulness (even with gravitational effects included).  For 
example, in the linearised problem (\ref{svelf}) at finite wavenumber, the 
leading branch of the spectrum merges with the next branch, the gravest 
shear mode $u_1=\sin \pi\zeta/2$, at a wavenumber $k\approx 2$ to form a 
pair of oscillatory decaying modes.  Such oscillations are the remnants, 
under the strong viscous dissipation in thin films, of the waves which 
surface tension can support.  In many applications, such decaying waves seem 
important, for example see the review by Chang\cite{Chang94}.  However, the 
model~(\ref{svenm}) {\em cannot} describe the necessary oscillations 
because it has only one component and is only first-order in time.  This 
appears at first sight to be a strong limitation to the practical 
usefulness of centre manifold techniques in applications.  But with some 
imagination we can modify the governing equations so that a centre manifold 
model is formed based on the {\em two\/} leading branches of the spectrum 
\cite{Roberts94c}.  Such a model has much wider applicability because it is 
a much improved description at finite wavenumber and it resolves shorter 
transients in time.

\section{Conclusion}

In summary, the proposed algorithm for the computer algebra derivation of 
low-dimensional models of dynamical systems is relatively:
\begin{itemize}
\item simple to implement, because the computation of the residual is via a 
{\em direct\/} coding of the governing differential equations;
	
\item reliable, because it relies upon the actual residual going to 
zero---any error is picked up by a lack of convergence;
	
\item flexible, because it removes the explicit tyranny of primitive 
approaches, such as the method of multiple scales, in forcing highly 
specific scalings upon the parameters and variables in the model---instead 
centre manifold theory assures the model is accurate to the order of 
accuracy of the residual.
\end{itemize}

\paragraph{Acknowledgements:} I thank Val\'ery Roy for stimulating 
discussions during the preparation of this work.

\end{document}